\documentclass[referee,useAMS,usenatbib,usegraphicx]{mn2e}

\newcommand{\xmm}{\it XMM-Newton}
\newcommand{\suzaku}{\it Suzaku}

\def\1705{4U~1705--44}

\usepackage{subfigure}
\bibpunct{(}{)}{;}{a}{}{,}

\title[\suzaku\ broad--band spectrum of \1705]
{\suzaku\ broad band spectrum of \1705: Probing the Reflection 
component in the hard state}

\author[T. Di Salvo et al.]{T. Di Salvo$^{1}$\thanks{E-mail:tiziana.disalvo@unipa.it},
R. Iaria$^{1}$, M. Matranga$^{1}$, L. Burderi$^{2}$, A. D'A\'\i$^{1}$,  
\vspace*{0.3cm}
\\ {\LARGE \textup{E. Egron$^{3}$, A. Papitto$^{4}$, A. Riggio$^{2}$, N.~R. Robba$^{1}$, Y. Ueda$^{5}$} }
\vspace*{0.1cm}
\\
$^1$Dipartimento di Fisica e Chimica,
Universit\`a degli Studi di Palermo, via Archirafi 36 - 90123 Palermo, Italy\\
$^2$Universit\`a degli Studi di Cagliari, Dipartimento
di Fisica, SP Monserrato-Sestu, KM 0.7, 09042 Monserrato, Italy \\
$^{3}$INAF -- Osservatorio Astronomico di Cagliari, via della Scienza 5, 09047 
Selargius (CA), Italy   \\
$^{4}$Institut de Ci\'encies de l'Espai (IEEC-CSIC), Campus UAB, Fac. de Ci\'encies, 
Torre C5, parell, 2a planta, 08193, Barcelona, Spain    \\
$^{5}$Department of Astronomy, Kyoto University, Kitashirakawa-Oiwake-cho, Sakyo-ku, 
Kyoto 606-8502, Japan   } 

\begin{document}

\date{}

\maketitle

\begin{abstract}
Iron emission lines at $6.4 - 6.97$ keV, identified with K$\alpha$ 
radiative transitions, are among the strongest discrete features in the 
X-ray band. These are one of the most powerful probes to infer the
properties of the plasma in the innermost part of the accretion disk
around a compact object.  In this paper we present a recent {\suzaku} 
observation, 100--ks effective exposure, of the atoll source and X-ray 
burster {\1705}, where we clearly detect signatures of a reflection
component which is distorted by the high--velocity motion in the accretion 
disk. The reflection component consists of a broad iron line at about 
6.4~keV and a Compton bump at high X-ray energies, around 20 keV.
All these features are consistently fitted with a reflection model, 
and we find that in the hard state the smearing parameters are remarkably
similar to those found in a previous {\xmm} observation performed in the
soft state. In particular, we find that the inner disk radius is 
$R_{in} = 17 \pm 5\; R_g$ (where $R_g$ is the Gravitational radius, 
$GM/c^2$), the emissivity dependence from the disk radius is 
$r^{-2.5 \pm 0.5}$, the inclination angle with respect to the 
line of sight is $i = 43^\circ \pm 5^\circ$, and the outer radius of
the emitting region in the disk is $R_{out} > 200 \; R_g$. 
We note that the accretion disk does not appear to be truncated at
large radii, although the source is in a hard state at $\sim 3\%$ of
the Eddington luminosity for a neutron star. 
We also find evidence of a broad emission line at low energies, at $3.03
\pm 0.03$ keV, compatible with emission from mildly ionized Argon 
(Ar XVI-XVII). Argon transitions are not included in the self--consistent 
reflection models that we used and we therefore added an extra 
component to our model to fit this feature. The low energy 
line appears compatible with being smeared by the same inner disk 
parameters found for the reflection component.
\end{abstract}


\section{Introduction}

Neutron Star Low Mass X-ray Binaries (hereafter NS LMXBs) are binary systems 
in which a weakly magnetic NS accretes matter from a low mass ($< 1 M_\odot$) 
companion star via Roche-Lobe overflow. In these systems the accretion disk 
can approach the compact object, as testified by the very fast time 
variability observed up to kHz frequencies \citep[see][as a review]{vanderKlis:06}. 
Broad emission lines (FWHM up to $\sim 1$ keV) at energies in the range 
6.4 -- 6.97 keV are often observed in the spectra of NS LMXBs
\citep[e.g.][]{Di_Salvo.etal:05, Piraino.etal:07, Bhatta.etal:07, Cackett.etal:08, 
Pandel.etal:08, Di_Salvo.etal:09, D_Ai.etal:09, Iaria.etal:09, Papitto.etal:09, 
Cackett.etal:09, Shaposhnikov.etal:09, Papitto.etal:10, 
Egron.etal:11, Piraino.etal:12, Papitto.etal:13, 
Miller.etal:13}. These lines are identified with the K$\alpha$ radiative
transitions of iron at different ionization states.  These features
are powerful tools to investigate the structure of the accretion flow
close to the central source; in particular, important information can
be obtained from the detailed spectroscopy of the line profile, since
it is determined by the ionization state, geometry and velocity field
of the reprocessing plasma.
In fact, the broad iron line observed in NS LMXBs is thought to originate 
from reflection of the primary X-ray continuum off the inner accretion disk 
and the width of the line is expected to be a signature of the Keplerian 
motion of matter in the inner accretion disk at (mildly) relativistic 
velocities. In this model, the combination of Doppler effects from the high 
orbital velocities and Special and General relativistic effects arising from
the strong gravity in the vicinity of the NS smears and shifts the reflected 
features. As a consequence, the line will have a characteristically broad 
and asymmetric profile, the detailed shape of which depends on the 
inclination and on how deep the accretion disk extends into the NS 
gravitational potential \citep[e.g.][]{Fabian.etal:89, Matt.etal:92}.

If the origin of this line is from disk reprocessing, one would also 
expect the presence of other discrete features (such as emission lines and 
absorption edges from the most abundant elements) and a bump between 20 
and 40 keV due to direct Compton scattering of the primary spectrum by 
the electrons in the disk. Indeed this reflection bump has been observed 
in the spectra of some NS LMXBs \citep[see e.g.][]{Barret.etal:00, 
Piraino.etal:99, Yoshida.etal:93, Fiocchi.etal:07, Egron.etal:13, 
Miller.etal:13}, 
usually with reflection amplitudes (defined in terms of the solid angle
$\Omega/2\pi$ subtended by the reflector as seen from the corona) lower
than 0.3. In some cases an anti-correlation has been claimed between the
photon index of the primary spectrum and the reflection amplitude of the
reprocessed component \citep{Zdziarski.etal:99, Barret.etal:00, 
Piraino.etal:99}, the same observed in Seyfert galaxies and galactic Black 
Hole (hereafter BH) candidates. This is probably caused by 
variations in the position of the inner rim of the disk. 

The disk origin of the iron line in NS LMXBs is, however, debated in
literature because of the brightness of these sources, which may cause 
photon pile-up and systematics in CCD spectra \citep{Ng.etal:10},
making the detection of any asymmetry in the line profile somewhat 
controversial. However, a large number of simulations using a statistical 
model of photon pile-up to assess its impacts on relativistic disk line 
and continuum spectra suggest that severe photon pile-up acts to falsely 
narrow emission lines, leading to falsely large disk radii 
\citep{Miller.etal:10}. These simulations also indicate that relativistic 
disk spectroscopy is generally robust against pile-up when this effect is 
modest.
Moreover, several authors \citep[e.g.][]{Cackett.etal:12, Egron.etal:13} have shown 
that the CCD-based spectra from {\suzaku} and {\xmm} are compatible with  
gas-based spectra from EXOSAT, BeppoSAX, and RXTE, demonstrating 
that the broad profiles seen are intrinsic to the line and not due to 
instrumental issues. They also report that a few BeppoSAX observations 
show evidence for asymmetric lines, with a relativistic diskline model 
providing a significantly better fit than a Gaussian line 
\citep[see also][]{Piraino.etal:99}. 

Nevertheless alternatives have been proposed to explain the
profiles of these features. 
In particular, \citet{Ng.etal:10} propose that Compton broadening may
be sufficient to explain the large width of the line. However, when
self-consistently included in the fit, Compton broadening alone
appears to be insufficient to explain the observed line profile
\citep[see e.g.][]{Reis.etal:09, Egron.etal:13, Sanna.etal:13}.
Also, \citet{Cackett.Miller:13} have explored the observational signatures 
expected from broadening in a wind. In this case the iron line width 
should increase with increasing the column density of the absorber 
(due to an increase in the number of scatterings). They show that there 
is no significant correlation between line width and column density,
favoring an inner disk origin for the line broadening rather than scattering 
in a wind.

\1705 is a well-studied atoll source \citep[see][]{Hasinger.Klis:89, Olive.etal:03}, 
which also shows type-I X-ray bursts. 
Similarly to X-ray binaries containing BHs, this source regularly shows 
state transitions: from a high/soft state, where the X-ray 
spectrum is dominated by soft spectral components with typical temperatures 
less than a few keV, to low/hard states where the X-ray spectrum is
dominated by a hard thermal Comptonisation \citep[e.g.][]{Barret.Olive:02, 
Piraino.etal:07}. The presence of broad discrete 
features in this source has been often reported in literature.
A broad iron line was observed, in the soft and/or hard state, with 
moderate/high spectral resolution by the Chandra/HETG \citep{Di_Salvo.etal:05}, 
BeppoSAX \citep{Piraino.etal:07}, Suzaku \citep{Reis.etal:09},
and XMM/pn \citep{Di_Salvo.etal:09, D_Ai.etal:10, Egron.etal:13}.
The XMM observation, taken in August 2008 during a soft state (45 ks effective
exposure time), showed one of the highest signal-to-noise ratio iron line profile 
ever observed in a NS LMXB. The line profile is clearly broad 
and could be fitted equally well with a relativistic line 
profile, such as {\it diskline} \citep{Fabian.etal:89} or {\it relline} 
\citep{Dauser.etal:10}, or with self-consistent reflection models, such as 
{\it reflionx} \citep{Ross.Fabian:05}, {\it refbb} \citep{Ballantyne:04}, 
and {\it xillver} \citep{Garcia.Kallman:10}. 
All these models gave parameters of the inner disk with unprecedented 
precision and all compatible with each other within the small statistical 
uncertainties \citep[see][]{Di_Salvo.etal:09, D_Ai.etal:10, Egron.etal:13}. 
The line is identified with the K$\alpha$ transition of 
highly ionised iron, Fe XXV; the inner disk radius is $R_{in} = 14 \pm 2$ R$_g$ 
(where R$_g$ is the Gravitational radius, $G M / c^2$), the emissivity index 
of the disk is $-2.27\pm0.08$ (compatible with a disk illuminated by a 
central source), the inclination angle with respect to the line
of sight is $i = 39 \pm 1$ degrees. 
This, together with the presence of other low-energy features from 
S XVI, Ar XVIII, Ca XIX and a smeared iron edge at 8.4 keV, which all are 
compatible with being smeared with the same inner disk parameters, 
makes \1705 the best source for a detailed spectroscopic study, in 
order to address the disk origin of the observed iron line and 
of the whole reflection component.

In this paper we present a high statistics, 100-ks effective exposure, {\suzaku} 
observation of {\1705} during a hard state: these data allow us a detailed 
study of the reflection features and the fit, with a self-consistent 
reflection model, of both the iron line profile and the associated Compton 
reflection bump at energies above 10 keV. 
In this spectrum, which includes hard-band data (up to 200 keV), 
the overall fractional amount of reflection is well determined by fitting the
Compton bump. We can therefore test whether the observed iron line is consistent 
with this fractional amount of reflection. 
In this way we confirm independently (fitting a different spectral state and
using different instruments) the inner disk parameters already obtained with 
{\xmm} in the soft state.

\section{Observations}

Suzaku \citep{Mitsuda.etal:07} observed {\1705} on 2012 March 27 as the result
of a Target of Opportunity (ToO) program during a hard state for a total 
observing time of 250 ks, corresponding to an effective exposure time of 
about 100 ks because of observational gaps caused by Earth occultations 
along the low equatorial orbit of the {\suzaku} satellite.
Both the X-ray Imaging Spectrometers (XIS, 0.2-12 keV; \citet{Koyama.etal:07})
and the Hard X-ray Detector (HXD, 10-600 keV; \citet{Takahashi.etal:07})
instruments were used during these observations.
There are four XIS detectors, numbered as 0 to 3. XIS0, XIS2 and XIS3
all use front-illuminated CCDs and have very similar responses, while
XIS1 uses a back-illuminated CCD. At the time of this observation the
available CCDs were three due to the loss of the XIS2. The HXD instrument
consists of two types of collimated (non-imaging) detectors, 
the PIN diodes (10-70 keV) and the GSO scintillators (30-600 keV).

We reprocessed the data using the \emph{aepipeline} tool provided by
\emph{Suzaku FTOOLS version 20} updated with the latest calibration files
(2013 November).
As second step, in order to obtain a more accurate estimate of the
\emph{Suzaku} attitude, we have calculated a new attitude using the
free tool \emph{aeattcor.sl} created by J.~E. Davis. Then we have applied
the new attitude to XIS event files using the FTOOLS \emph{xiscoord}.
During the observation, XIS0, XIS1, and XIS3 were operated using the 1/4 
window option. The effective exposure time of each XIS CCD is 96.67 ks.
In order to estimate the pile-up in the XIS spectra we have used
the public available tool \emph{pile-estimate.sl} created by M.~A. Novak.
Using a circular region with radius equal to 105\arcsec, we have found that
in each XIS CCD, the pile-up fraction is $\sim 3\%$ at most. The pile-up
fraction is sufficiently small that we can neglect its effects on our spectral
fitting results. In fact, we have checked that spectral results do not change 
significantly if we exclude a central circle in the extraction region in order
to further reduce the pile-up fraction. Therefore, we have extracted the source 
and background spectra from a circular extraction region of radius 105\arcsec\ 
each, 
the background circle being centered close to the edge of the CCD, where no 
significant contaminating photons from the source were present. The response 
files of each XIS spectrum have been generated using the \emph{xisrmfgen} and 
\emph{xisarfgen} tools.
Since the response of XIS0 and XIS3 are very similar, we have combined their
spectra and responses using the tool \emph{addascaspec}.

The PIN spectrum has been extracted using the tool \emph{hxdpinxbpi}.
Both the non X-ray and cosmic X-ray backgrounds are
taken into account. The non X-ray background (NXB) is calculated
from the background event files distributed by the HXD
team. The cosmic X-ray background (CXB) is from the model
by \citet{Boldt.etal:87}. The response files provided by the HXD team
are used.
We selected the HXD/PIN events in the energy range 12-30 keV and produced
the HXD/PIN background-subtracted light curve using the SUZAKU tool
{\tt hxd-pinxblc} and adopting the background event files distributed by
the HXD team. Since this light curve follows the XIS0 light curve, we
conclude that no contaminating flare was present in the data.
We have also extracted the GSO spectrum running the tool \emph{hxdgsoxbpi}.
For the background we have used the 'tuned' non X-ray background,
whereas for the response file we have used the latest version provided on
2011-06-01.

We have extracted the XIS0 light curves in the energy range $0.9-2.8$,
$2.8-10$ and $0.9-10$ keV, respectively (see Fig.~1). Nine type-I X-ray 
bursts are observed in the total 250-ks light curve. The source (persistent) 
count rate gradually increases by about 40\% during the observation. 
Since no changes are observed in the hardness ratio (given by the ratio of 
the source counts in the $2.8-10$ keV range to the source counts in the 
$0.9-2.8$ keV, see Fig.~1) we conclude that the X-ray spectral shape of the 
source does not change significantly during the observation.

\begin{figure}
\begin{center}
    \begin{minipage}{16cm}
      \includegraphics[width=6cm,angle=270]{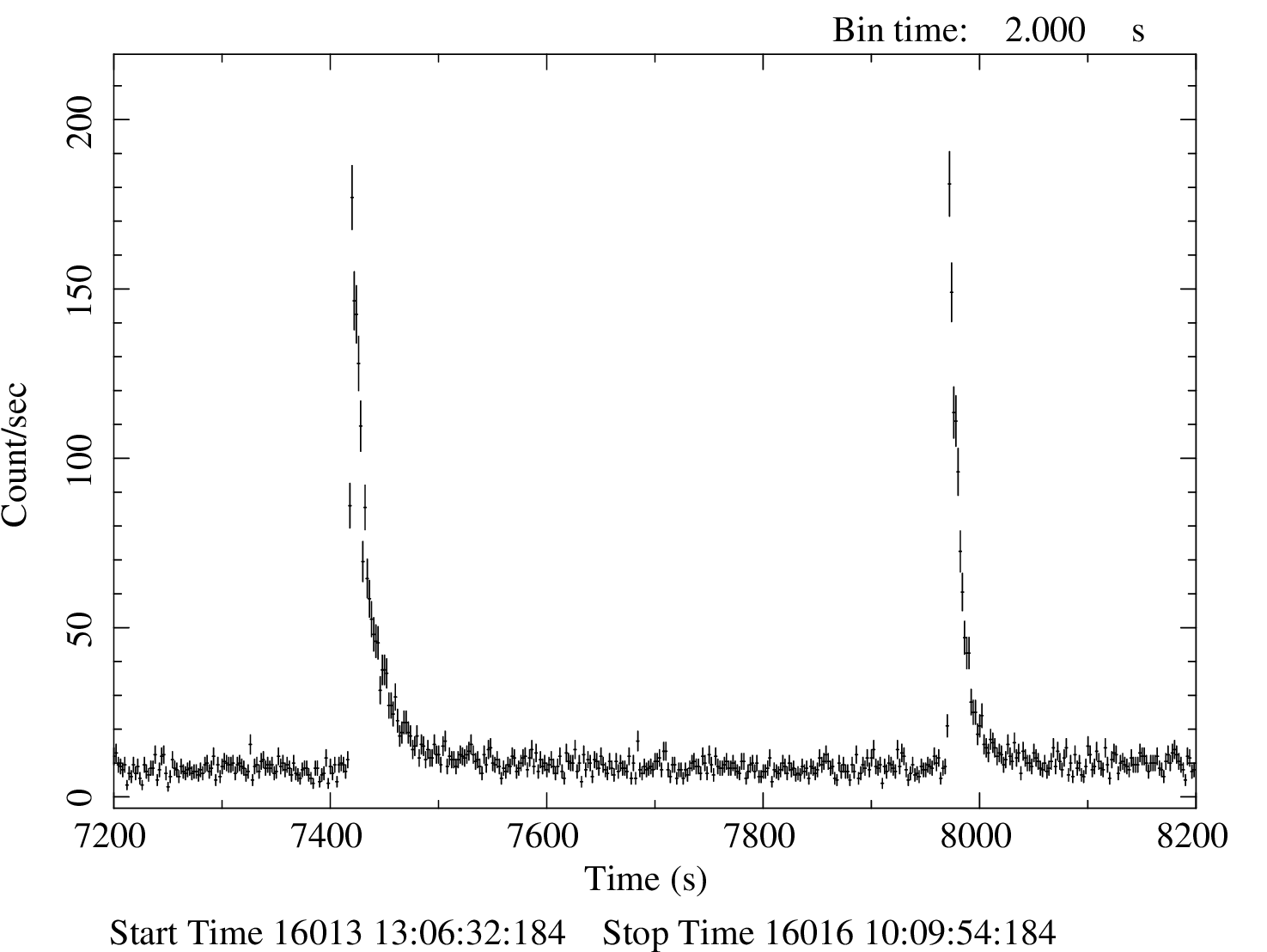}
      \includegraphics[width=6cm,angle=270]{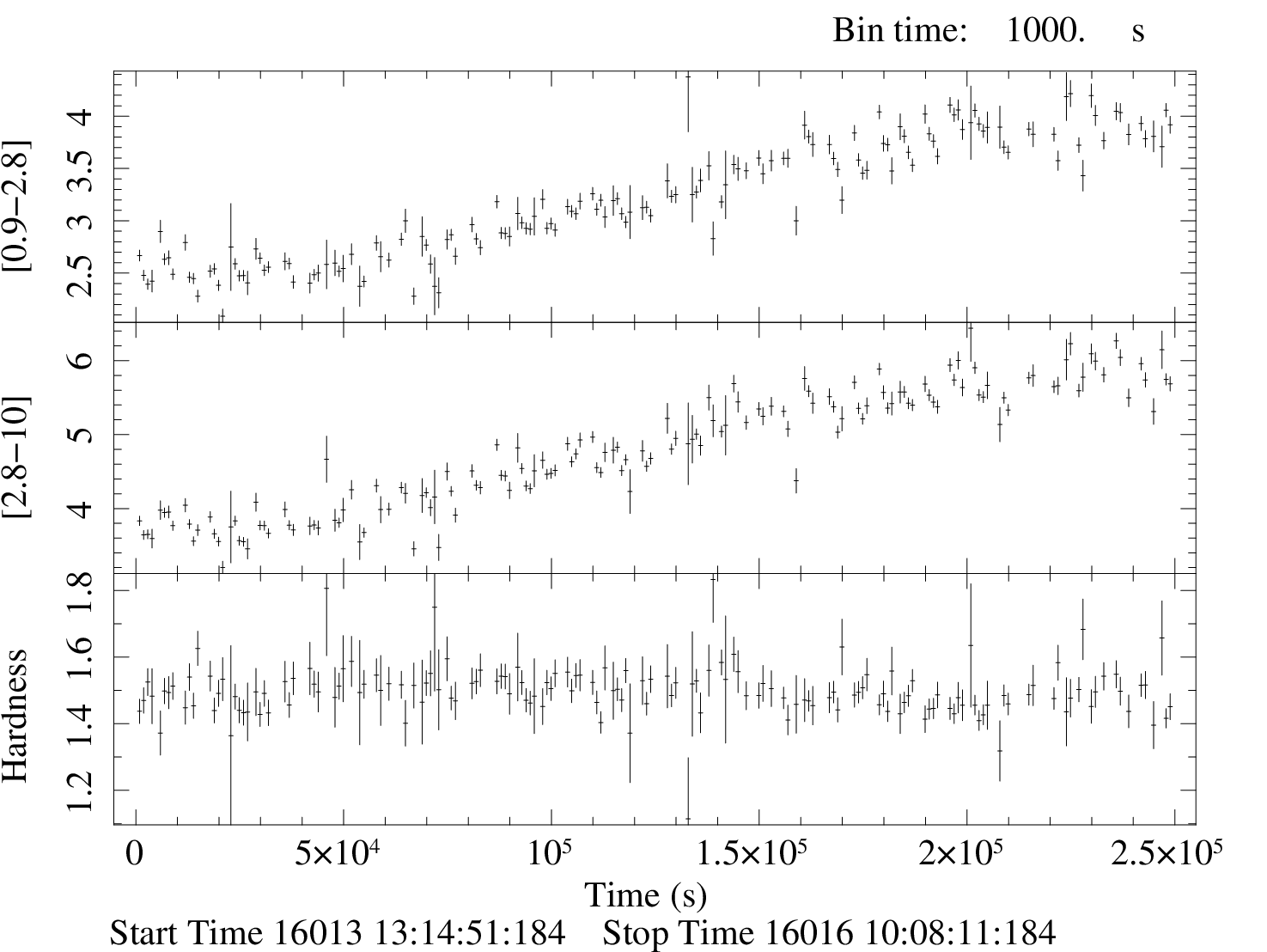}
    \caption{{\bf Left:} {\suzaku} XIS0 light curve in the energy range 
    0.9 - 10 keV showing two of the nine type-I bursts which occurred 
    during the 250-ks observation.
    {\bf Right:} {\suzaku} XIS0 light curves in the energy range $0.9-2.8$
    keV (top panel), $2.8-10$ keV (middle panel), and the corresponding 
    hardness ratio (bottom panel).}
    \label{fig2}
    \end{minipage}
\end{center}
\end{figure}

\section{Spectral Analysis and Results}

In order to extract spectra for the persistent emission, we have excluded 
the type-I bursts that occurred during the observation. In particular we have 
excluded approximately 100 s of data starting from the onset of each burst.
We adopt $0.7-11$ keV energy range for the XIS0+XIS3 (hereafter 
XIS03) and XIS1 spectra, $15-50$ keV energy range for the HXD/PIN 
spectrum and $50-200$ keV energy range for the HXD/GSO spectrum. 
We excluded the energy interval between 1.7 and 2.0 keV from the XIS03 
and XIS1 spectra because of the presence of systematic features associated 
with neutral silicon and neutral gold which give a mismatch between the 
two spectra. The XIS spectra were grouped by a factor 4 in order not to 
oversample too much the instrumental energy resolution. The HXD/PIN and 
HXD/GSO spectra were grouped in order to have at least 25 photons per 
energy channel. We fitted the spectra using XSPEC version 12.7.0.

We started to fit the continuum in the 0.7--200 keV energy range
with the typical model used for NS LMXBs of the atoll class,
which revealed to be the best fit continuum for this source too 
\citep[see e.g.][]{Di_Salvo.etal:09, Piraino.etal:07, Barret.Olive:02}. 
This model consists of a soft blackbody and a thermal Comptonised 
component, in this case modelled by {\tt nthComp} \citep{Zycki.etal:99}, 
modified at low energy by photoelectric absorption caused by neutral 
matter and modeled by {\tt phabs} in XSPEC. 
This continuum model gave, however, an unacceptable fit, corresponding to 
a $\chi^2 / dof = 2425.13 / 1511$, because of the presence of evident 
localised residuals at 2.5--3.5 keV, 6--9 keV and 15--30 keV. The
most prominent is a clear iron line profile at energies from 5 to
7 keV and an absorption feature at 7--8 keV (see Fig.~2).

\begin{figure}
\begin{center}
    \begin{minipage}{16cm}
      \includegraphics[width=11cm,angle=270]{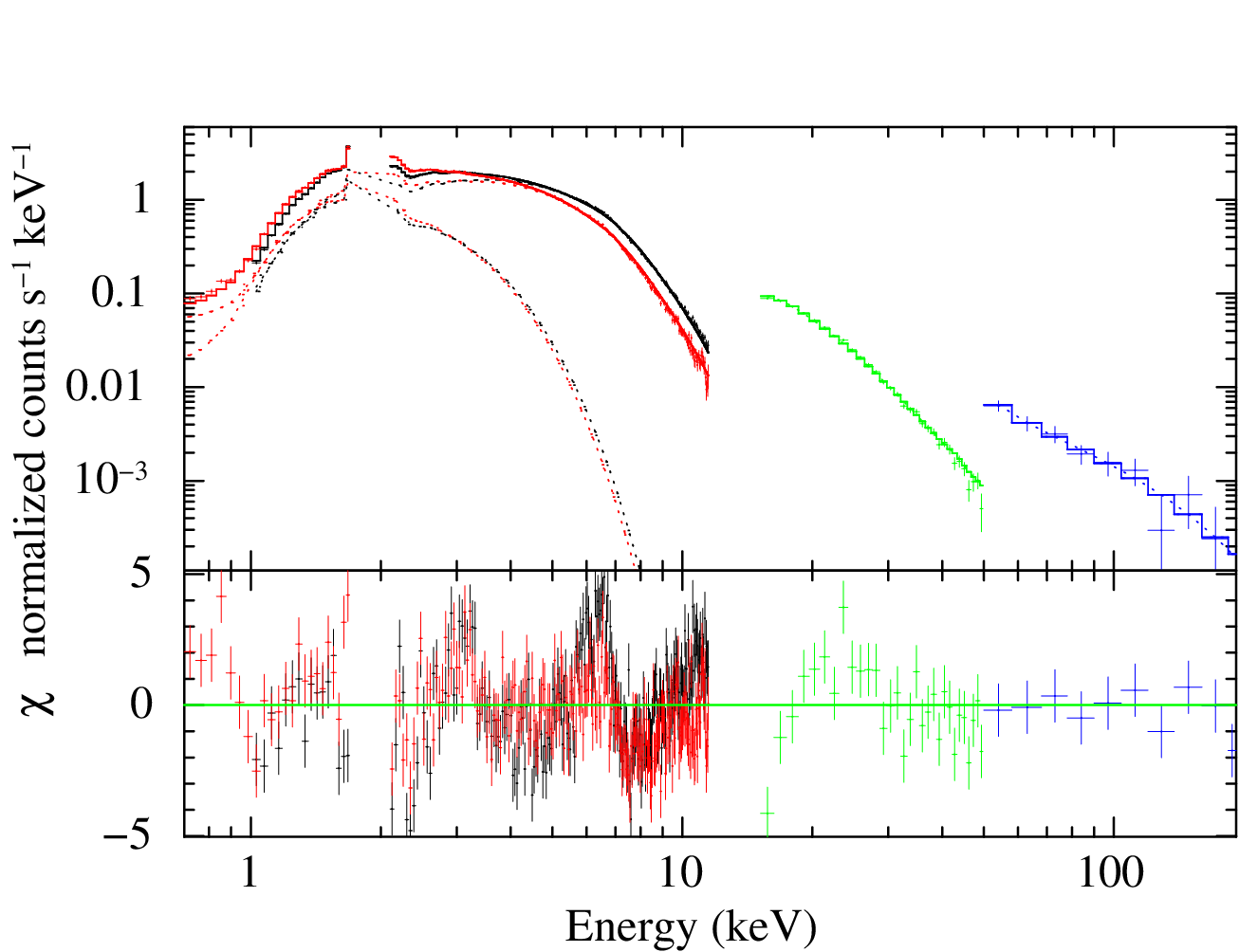}
    \caption{{\suzaku} data in the energy range 0.7 - 200 keV  (top) and residuals in 
    units of $\sigma$ with respect to the simpler phenomenological model (bottom) 
    of {\1705}. The model consists of a blackbody (dotted lines) and the Comptonization 
    component {\it nthComp}, both multiplied by  photoelectric absorption. }
    \label{fig2}
    \end{minipage}
\end{center}
\end{figure}

In order to fit these residuals, we first added to our continuum model 
the {\tt pexriv} component \citep{Magdziarz.etal:95} which takes into 
account the iron edge and Compton bump present in the residuals. Note
that pexriv does not self-consistently include any emission line.  
The photon index and the high energy cutoff of the pexriv component 
were fixed to those of the nthComp component. Here and in the following
we neglect any reflection of the soft (blackbody) component, which in
any case contributes to a small fraction of the total flux and most
of its flux is at soft energies (below $4-5$ keV).
The addition of this component gave a significant improvement of the 
fit reducing the $\chi^2 / dof$ to the value $2009.67 / 1509$ 
($\Delta \chi^2 = 415$ for the addition of two parameters).
Therefore, the presence of the Compton hump and the iron edge are detected 
with high statistical significance (an F-test would give a probability of 
chance improvement of the fit negligibly small, $\sim 2.7 \times 10^{-62}$).
We then added a Gaussian line at 6.4 keV obtaining again an improvement 
of the fit. As a first step we decided to fix the energy of the line
at 6.4 keV because, otherwise, it tends to get broad (Gaussian sigma about
1 keV) and its energy tends to decrease to 6.2 keV. This fit gives a 
$\chi^2 / dof = 1956.50 / 1507$ ($\Delta \chi^2 = 53$ for the addition of 
two parameters). 
This line can be identified with fluorescence from neutral iron. The addition 
of another Gaussian at about 3 keV again improves the fit, giving
$\chi^2 / dof = 1829.89 / 1504$ ($\Delta \chi^2 = 127$ for the addition 
of three parameters). This line can be identified with emission from mildly 
ionized Argon, Ar XVI--XVII.
In order to check whether the iron line energy was stable or not, we let 
the iron line energy free to vary obtaining $\chi^2 / dof = 1818 / 1503$. 
We also tried to substitute the Gaussian line at $\sim 6$ keV with a diskline.
In this case we had to fix all the smearing parameters but the inner radius
of the disk. This did result in a slight improvement 
of the fit, since we get $\chi^2 / dof = 1803 / 1503$ fixing the emissivity
index at $-2.4$, the outer radius at 400 R$_g$, the inclination angle at 40 deg.
Also in this case the centroid energy of the line remained at $6.1-6.3$ keV,
but we get an upper limit to the inner radius of the disk of 25 R$_g$.
On the other hand, we get a more significant improvement of the fit if we 
add a (mild) relativistic smearing to the whole reflection component (i.e.\ 
the emission lines at $\sim 3$ and $\sim 6$ keV and the pexriv component) 
convolving all these three components with {\tt rdblur}, the kernel of 
{\tt diskline}.
In this case we get $\chi^2 / dof = 1783 / 1501$ ($\Delta \chi^2 = 35$ for 
the addition of three parameters, corresponding to an F-test probability
of chance improvement of $\sim 4 \times 10^{-7}$).
The results of these phenomenological models are reported in Table~1.

\begin{table}
\caption{The best fit parameters of the spectral fitting of the {\suzaku} spectrum 
of {\1705} in the $0.7-200$ keV energy range with phenomenological 
models. The blackbody luminosity is given in units of $L_{35}/D_{10}^2$,
where $L_{35}$ is the bolometric luminosity in units of $10^{35}$~ergs/s and 
$D_{10}$ the distance to the source in units of 10~kpc. The blackbody radius
is calculated in the hypothesis of spherical emission and for a distance of
7.4~kpc. 
Fluxes in the nthComp and pexriv components are calculated in the $1-16$ keV 
range, while total flux is calculated in the $1-10$ keV band. 
Uncertainties are given at $90\%$ confidence level.
}
\vskip 0.5cm
\begin{tabular}{|llcccc|}
\hline
Component & Parameter & Basic Model & Pexriv  & Pexiriv + 2Gauss  &  Smeared (Pex + 2Gaus) \\
\hline
phabs & $N_H$ ($\times 10^{22}$ cm$^{-2}$) & $1.899 \pm 0.025$ & $2.13 \pm 0.05$  &  $2.02 \pm 0.05$ 
                                           & $2.04 \pm 0.06$ \\
bbody & $kT_{BB}$ (keV)          & $0.503 \pm 0.015$ &  $0.265^{+0.004}_{-0.026}$  &  $0.39 \pm 0.04$ 
                                           & $0.35 \pm 0.04$ \\
bbody & L$_{BB}$ ($L_{35} / D_{10}^2$) / Norm   & $8.8 \pm 0.6$ & $3.6 \pm 0.8$  &  $5.3 \pm 1.3$  
                                           & $4.0 \pm 0.6$ \\
bbody & R$_{BB}$ (km)  & $7.6 \pm 0.5$  &  $17.6 \pm 4.4$  &   $9.8 \pm 2.3$ & $10.6 \pm 2.5$ \\
nthComp & $kT_{seed}$ (keV)		   & $0.90 \pm 0.04$  & $0.569 \pm 0.014$  &  $0.69 \pm 0.06$ 
                                           & $0.64 \pm 0.04$ \\
nthComp& $\Gamma$                          & $2.05 \pm 0.03$ &  $2.081 \pm 0.018$  & $2.05 \pm 0.04$ 
                                           & $2.08 \pm 0.03$ \\
nthComp & $kT_e$ (keV)			   & $101^{+100}_{-74}$ & $63.2^{+12}_{-2.4}$  
                                           & $80.7^{+59}_{-9.9}$ & $89^{+28}_{-20}$  \\
nthComp & Flux ($10^{-10}$ ergs cm$^{-2}$ s$^{-1}$) &    $4.73$  &    $4.67$ &  $4.66$ 
                                           & $4.61$ \\

pexriv & $\xi$ (erg cm s$^{-1}$)             & -- & $< 1$ & $< 1$ & $< 1$  \\  
pexriv & Incl (deg)       & -- &  40 (fixed) &   40 (fixed) & 40 (fixed) \\  
pexriv & Flux ($10^{-10}$ ergs cm$^{-2}$ s$^{-1}$) &      &    $0.9$ &  $0.56$ & $0.73$ \\

gauss & $E_{line}$ (keV) & --  & --   &  $3.00 \pm 0.04$  &  $3.02 \pm 0.04$  \\
gauss & $\sigma_{line}$  (keV) & --  &  --  &  $0.28 \pm 0.03$ & --  \\
gauss & $I_{line}$ ($\times 10^{-4}$ ph cm$^{-2}$ s$^{-1}$) & --  & --   &  $3.3 \pm 0.9$ 
							    & $2.4 \pm 0.5$ \\
gauss & EqW (eV)         & --  & --  &  $24.0 \pm 7.6$ & $18.3 \pm 3.5$  \\
gauss & $E_{Fe}$ (keV) & --  & --   &   $6.21 \pm 0.08$ & $6.27 \pm 0.06$  \\
gauss & $\sigma_{Fe}$ (keV) & --  &  --  &  $0.46 \pm 0.07$ & --  \\
gauss & $I_{line}$ ($\times 10^{-4}$ ph cm$^{-2}$ s$^{-1}$) & --  & --   &  $1.58^{+1.2}_{-0.19}$ 
							    & $1.9^{+0.7}_{-0.4}$ \\
gauss & EqW (eV)         & --  & --  &  $33 \pm 17$  &  $52 \pm 21$  \\

rdblur & $Betor$         & --  & --  &  --  &  $-2.2 \pm 0.5$  \\  
rdblur & $R_{in}$ ($G M / c^2$)   & -- & --  &  --  &  $ < 29 $  \\  
rdblur & $R_{out}$ ($G M / c^2$)  & -- & --  &  --  &  $360^{+360}_{-160}$ \\  
rdblur & Incl (deg)       & -- & -- & -- &  $54^{+17}_{-9} $ \\  

\hline
total & Flux ($10^{-10}$ ergs cm$^{-2}$ s$^{-1}$)  & $3.3 \pm 0.7$  & $3.34 \pm 0.02$   
						   &  $3.34 \pm 0.03$ & $3.34 \pm 0.03$ \\
total &  $\chi^2$ (dof)	   & $2425~(1511)$ &  $1990~(1509)$ &  $1818~(1503)$ & $1783~(1501)$  \\
\hline
\end{tabular}
\end{table}

In order to fit the high-quality {\suzaku} spectrum with more consistent physical 
models, we substitute the pexriv + Fe line components with the self-consistent 
reflection model {\tt reflionx} \citep{Ross.Fabian:05}, modified by a relativistic 
blurring component (again modeled with rdblur) to consider 
the relativistic and/or Doppler effects produced by the motion in the inner disk 
close to the compact object. In this model, emission lines from the most abundant 
elements or ions are also self-consistently calculated.
In the reflionx model the emergent (reflected) spectrum is calculated for an 
optically-thick atmosphere (such as the surface of an accretion disk) of 
constant density illuminated by radiation with a power-law spectrum, whose 
photon index is fixed to that of the nthComp component, and a 
high-energy exponential cutoff with e-folding energy fixed at 300 keV. In order
to take into account the high-energy cutoff in the illuminating spectrum, we have
multiplied the reflionx spectrum by a high energy cutoff with the e-folding 
energy fixed to the value of the e-folding energy of the primary (nthComp) 
component.
Since the reflionx model does not take into account transitions from Ar and Ca, we 
fitted the emission line at $\sim 3$ keV with a diskline, fixing all the smearing 
parameters to those used for the reflionx component. 
The continuum emission is fitted with the same model as before, and we used for the
soft component alternatively the bbody or the diskbb model. The results of these 
fits are shown in Table~2. We also checked the possibility of iron overabundance
with respect to cosmic abundances \citep[as claimed by][for the soft state]{Egron.etal:13}, 
fixing the iron abundance alternatively to the cosmic value and twice 
the cosmic value. We find a slightly better fit when we fix the iron abundance to 
the cosmic value and we use a blackbody to fit the soft thermal component (see Table~2).
The total $0.5-200$ keV luminosity of the source during the {\suzaku} observation was
$6.15Ê\times 10^{36}$ ergs/s assuming a distance to the source of 7.4 kpc.

\begin{table}
\caption{The best fit parameters of the spectral fitting of the {\suzaku} spectrum 
of {\1705} in the $0.7-200$ keV energy range with the self-consistent reflection 
model {\tt reflionx}. The blackbody luminosity is given in units of $L_{35}/D_{10}^2$,
where $L_{35}$ is the bolometric luminosity in units of $10^{35}$~ergs/s and 
$D_{10}$ the distance to the source in units of 10~kpc. The blackbody radius
is calculated in the hypothesis of spherical emission and for a distance of
7.4~kpc. The disk blackbody normalization is given by $(R_{in} (km)/D_{10})^2
\cos i$, where $i$ is the inclination angle of the binary system. The disk
blackbody inner radius R$_{in}$ (km) is calculated for an inclination angle
of $40^\circ$. 
Flux is calculated in the $1-10$ keV band. Uncertainties are given at 
$90\%$ confidence level.
}
\vskip 0.5cm
\begin{tabular}{|lcccc|}
\hline
Parameter & BBODY [Fe = 1] & DISKBB [Fe = 1] & BBODY [Fe = 2] & DISKBB [Fe = 2]  \\
\hline
$N_H$ ($\times 10^{22}$ cm$^{-2}$) & $2.11 \pm 0.04$ & $2.27 \pm 0.04$
& $2.09 \pm 0.04$ & $2.25 \pm 0.04$ \\
$kT_{BB}$ (keV)                   & $0.38 \pm 0.03$ & $0.53 \pm 0.07$
& $0.38 \pm 0.03$ & $0.52 \pm 0.08$  \\
L$_{BB}$ ($L_{35} / D_{10}^2$) / Norm   & $5.5 \pm 0.7$ & $58 \pm 22$
& $4.9 \pm 0.5$ & $61^{+42}_{-18}$  \\
R$_{BB}$ (km) / R$_{in}$ (km) & $10.6 \pm 1.8$ & $6.4 \pm 1.2$  
& $10.0 \pm 1.7$ & $6.6^{+2.3}_{-1.0}$  \\
$kT_{seed}$ (keV)		   & $0.68 \pm 0.03$ & $0.70 \pm 0.06$
& $0.66 \pm 0.03$ & $0.67 \pm 0.06$  \\
$\Gamma$                           & $2.01 \pm 0.02$ & $2.01 \pm 0.02$
& $1.95 \pm 0.01$  & $1.95 \pm 0.01$ \\
$kT_e$ (keV)			   & $47^{+19}_{-11}$ & $47^{+17}_{-10}$
& $27 \pm 5$  & $27 \pm 5$ \\
$\xi$ (erg cm s$^{-1}$)             & $< 13$ & $< 19$ & $< 24$  & $22 \pm 4$ \\  
$Betor$          & $-2.5 \pm 0.5$ & $-2.5 \pm 0.5$
& $-2.4 \pm 0.4$   & $-2.5 \pm 0.5$   \\  
$R_{in}$ ($G M / c^2$)  & $17^{+4}_{-6}$ & $16^{+4}_{-7}$
& $16^{+4}_{-5}$  & $16^{+4}_{-7}$ \\  
$R_{out}$ ($G M / c^2$) & $370^{+8000}_{-180}$ & $370^{+370}_{-170}$
& $330^{+8000}_{-200}$  & $350^{+350}_{-170}$ \\  
Incl (deg)       & $43 \pm 5$ & $42 \pm 5$
& $41 \pm 4$  & $42 \pm 4$ \\  
$E_{line}$ (keV) & $3.03 \pm 0.03$ & $3.03 \pm 0.03$
& $3.04 \pm 0.03$   & $3.04 \pm 0.03$ \\
$I_{line}$ ($\times 10^{-4}$ ph cm$^{-2}$ s$^{-1}$) & $2.4 \pm 0.4$ & $2.3 \pm 0.4$
& $2.4 \pm 0.4$  & $2.3 \pm 0.4$ \\
EqW (eV)         & $18.2 \pm 3.6$  & $17.0 \pm 3.4$
& $18.6 \pm 3.5$  & $17.5 \pm 3.7$ \\
\hline
Flux ($10^{-10}$ ergs cm$^{-2}$ s$^{-1}$)  & $3.34 \pm 0.04$ & $3.34 \pm 0.15$
& $3.34 \pm 0.04$  & $3.34 \pm 0.14$ \\
total $\chi^2$ (dof)		   & $1831~(1503)$ & $1841~(1503)$
& $1846~(1503)$  & $1852~(1503)$ \\
\hline
\end{tabular}
\end{table}

\begin{figure}
\begin{center}
    \begin{minipage}{16cm}
      \includegraphics[width=6cm,angle=270]{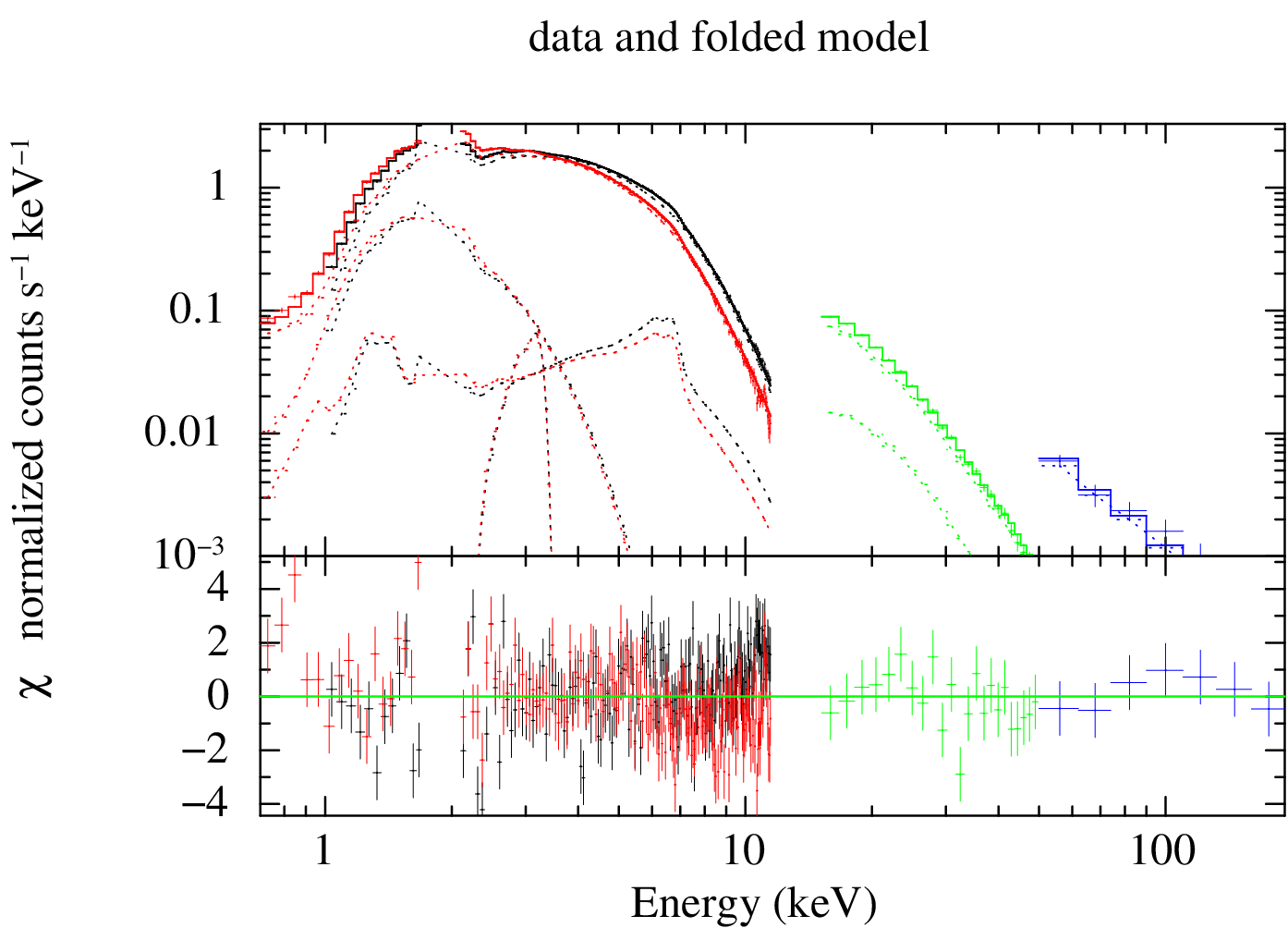}
      \includegraphics[width=6cm,angle=270]{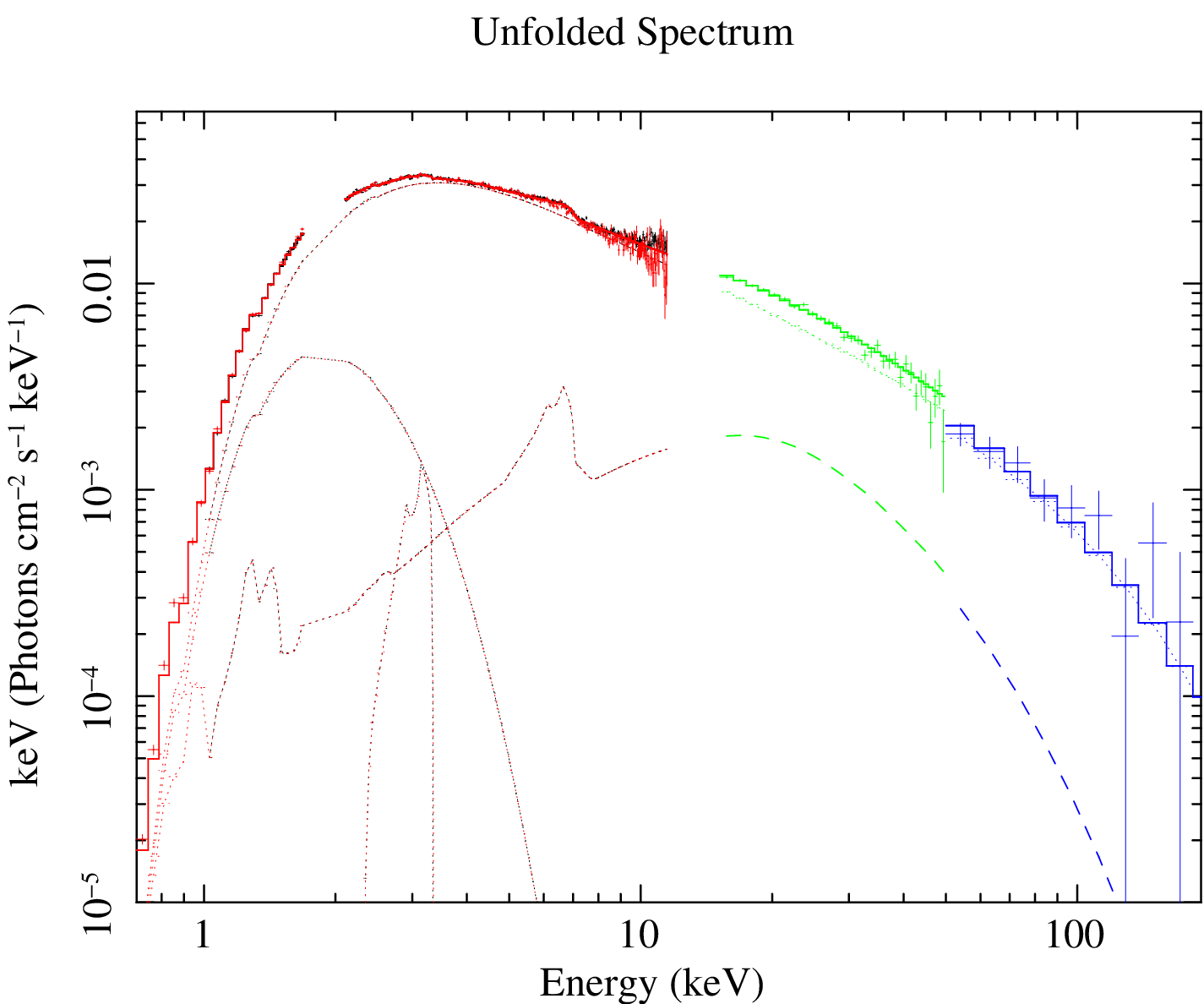}
    \caption{{\bf Left:} {\suzaku} data in the energy range 0.7 - 200 keV (top) 
    and residuals in units of $\sigma$ with respect to the best-fit model (bottom) 
    of {\1705} (see Table~2, first column). 
    {\bf Right:} {\suzaku} unfolded spectrum in the energy range 0.7 - 200 keV with
    respect to the best-fit model shown in the first column of Table~2. The model
    components are also shown. From the left to the right we see the blackbody
    component, the emission line at $\sim 3$~keV (smeared with the same smearing 
    parameters used for the reflection component), the smeared reflection component
    modeled by {\it reflionx}. The main Comptonization component and the total
    model are plotted on top of the data. }
    \label{fig3}
    \end{minipage}
\end{center}
\end{figure}

Finally, we fitted the reflection component in the {\suzaku} spectrum of \1705 
with the convolution model {\tt rfxconv} \citep{Kolehmainen.etal:11}, which has the
advantage to take into account the exact shape of the illuminating continuum.
For sake of completeness we also tried the self-consistent reflection model
{\tt relxill} by \citet{Garcia.etal:14}, whose novelty is that for each point 
on the disk the proper reflection spectrum is chosen for each relativistically 
calculated emission angle. In the relxill model we fixed the photon index and
e-folding energy of the reflected spectrum to those of the nthComp component. 
Both these models include emission lines from the most 
abundant elements or ions, as in the case of {\tt reflionx}.
Again we obtain a good fit of the \1705 spectrum, with best-fit values of the
reflection and smearing parameters well in agreement with those obtained with
other reflection models (see Tab.~3, cf.\ with Tab.~1 and 2).

\begin{table}
\caption{The best fit parameters of the spectral fitting of the {\suzaku} spectrum 
of {\1705} in the $0.7-200$ keV energy range with the self-consistent reflection 
models {\tt rfxconv} and {\tt relxill}. The blackbody luminosity is given in units 
of $L_{35}/D_{10}^2$, where $L_{35}$ is the bolometric luminosity in units of 
$10^{35}$~ergs/s and $D_{10}$ the distance to the source in units of 10~kpc. 
The blackbody radius is calculated in the hypothesis of spherical emission and 
for a distance of 7.4~kpc. Flux is calculated in the $0.7-200$ keV band. 
Uncertainties are given at $90\%$ confidence level.
}
\vskip 0.5cm
\begin{tabular}{|lcc|}
\hline
Parameter & RFXCONV [Fe = 1]  & RELXILL [Fe = 1]  \\
\hline
$N_H$ ($\times 10^{22}$ cm$^{-2}$) & $2.05 \pm 0.04$ &  $2.06 \pm 0.05$  \\
$kT_{BB}$ (keV)                   & $0.37 \pm 0.03$  &  $0.36 \pm 0.04$    \\
L$_{BB}$ ($L_{35} / D_{10}^2$) / Norm   & $4.6 \pm 0.6$ & $4.1 \pm 0.7$ \\
R$_{BB}$ (km) / R$_{in}$ (km) & $10.2 \pm 1.8$  &  $10.3 \pm 2.5$  \\
$kT_{seed}$ (keV)		   & $0.67 \pm 0.03$ &  $0.63 \pm 0.04$  \\
$\Gamma$                           & $2.01 \pm 0.01$ &  $1.979 \pm 0.009$  \\
$kT_e$ (keV)			   & $34^{+5}_{-4}$  &  $43 \pm 9$  \\
$\xi$ (erg cm s$^{-1}$)            & $60^{+20}_{-30}$ & $12^{+9}_{-4}$ \\  
Refl Amplitude   & $0.34 \pm 0.04$ & $0.34$ (fixed) \\   
Refl Norm        & $-$ & $0.161 \pm 0.017$  \\   
$Betor$          & $-2.5 \pm 0.5$  & $-3.2^{+0.4}_{-0.2}$  \\  
$R_{in}$ ($G M / c^2$)  & $17^{+4}_{-7}$  &  $14 \pm 2$ \\  
$R_{out}$ ($G M / c^2$) & $> 202$  &  $260$ (fixed)  \\  
Incl (deg)       & $43 \pm 5$      &  $31.6^{+1.9}_{-1.5}$  \\  
$E_{line}$ (keV) & $3.04 \pm 0.04$ &  $3.10 \pm 0.03$  \\
$I_{line}$ ($\times 10^{-4}$ ph cm$^{-2}$ s$^{-1}$) & $2.5 \pm 0.4$  &  $2.2 \pm 0.4$ \\
EqW (eV)         & $19.3 \pm 0.3$  &  $17^{+3}_{-2}$ \\
\hline
Flux ($10^{-10}$ ergs cm$^{-2}$ s$^{-1}$)  & $9.0 \pm 0.3$  &  $9.33^{+0.17}_{-0.36}$  \\
total $\chi^2$ (dof)		   & $1832~(1503)$  &  $1818~(1504)$ \\
\hline
\end{tabular}
\end{table}

\section{Discussion}

Similar to BH X-ray Binaries, NS LMXBs show clear differences in the
spectral parameters during hard and soft states. Studying these differences
is important in order to address different geometry or physical properties
of the inner central emitting region in these two spectral states and in order
to understand what causes the spectral transition. In fact, there is no 
general consensus on the hot corona--accretion disk geometry in these two spectral
states. Theoretically the hard Comptonized component may come from either a 
patchy corona, possibly powered by magnetic flares \citep[e.g.][]{Beloborodov:99}
or the base of a centrally located jet \citep[e.g.][]{Markoff.etal:05}. 
In both the cases, the thin accretion disk may extend close to the last marginally
stable orbit or the NS surface. Alternatively the thin disk may be truncated 
at large distances from the compact object, with the central region replaced 
by an advection-dominated accretion flow (ADAF) where, at high accretion rates, 
material may condense to form an inner optically thick disk 
\citep[see e.g.][]{Esin.etal:97}. This hot, inner flow can also act as the 
launching site of the jet. In this respect, much information may come from 
the study of the so-called reflection component in different spectral states 
of a source \citep[see][as a review]{Done.etal:07}.

In this paper we have reported the results of the spectral analysis of a long
{\suzaku} observation of the LMXB of the atoll class \1705. This was the result
of a ToO program intended to observe the source during a hard state. 
{\suzaku} observed the source for a total of 250 ks yielding a net exposure 
time of 100 ks. During the observation 9 type-I bursts were observed. We 
present here the spectral analysis of the persistent emission, while we will 
discuss the characteristics of the observed type-I bursts elsewhere.
We have fitted the persistent emission spectrum in the broad band range
between 0.7 and 200 keV using the continuum model which gave the best fit
of previous high-quality spectra of this source obtained with RXTE, BeppoSAX,
{\xmm}, and Chandra \citep[see e.g.][and references therein]{Egron.etal:13}.
The continuum model consists of a soft component modeled by {\tt bbody}
and a Comptonization component modeled by {\tt nthComp}, both multiplied 
by the {\tt phabs} component which takes into account photoelectric 
absorption by neutral matter in the interstellar medium. A smeared reflection 
component was necessary to obtain an acceptable fit of the broad--band {\suzaku} 
spectrum.
This component was necessary to fit high-energy residuals above about 15 keV
and a broad iron edge at about $8-9$ keV -- the addition of the {\tt pexriv} 
component to fit these two features gave a significant improvement of the 
quality of the fit with a probability of chance improvement, calculated with 
an F-test, which resulted to be negligibly small. 

We detect other reflection
features, such as a broad emission line at about 6.4 keV from neutral iron and 
a broad emission line at about 3 keV that we tentatively identify with the
K$\alpha$ transition from Ar XVI-XVII. These two features could be well fitted
by broad {\tt Gaussians}. Note that this Ar line is quite strong with respect 
to the observed iron line. If we consider the product of the element abundances 
($\sim 4.7 \times 10^{-5}$ and $\sim 3.6 \times 10^{-6}$ for iron and argon, 
respectively) by the fluorescence yields (which can be calculated using the
empirical formula yield $= Z^4/(30^4+Z^4)$, where $Z$ is the atomic number),
we obtain $\sim 1.7 \times 10^{-5}$ and $\sim 4 \times 10^{-7}$ for iron and
argon, respectively. This means that the Ar line strength should be $\sim 2.4 
\%$ of Fe line strength. This in the hypothesis that all the atoms of these
elements are in the ionization state producing the line. Note also that the 
observed line may depend on the illuminating continuum at that energy. 
In the case of \1705 in the soft state, where Ar and Ca lines are clearly 
detected by {\xmm} together with an Fe line, we find small differences in the 
line intensities (a factor 2 at most) and higher differences in the line 
equivalent width (up to a factor 8). However, when the spectrum is fitted to 
a self-consistent reflection model, a suitably modified version of the 
{\tt xillver} model by \citet{Garcia.Kallman:10} which includes Ar and Ca 
transitions, we find that the Ar line is well fitted by the reflection model 
with a slight overabundance by a factor 1.8 with respect to Solar abundances
\citep[see][]{Egron.etal:13}. We conclude therefore that the
simple calculation above is merely an order of magnitude estimation and 
that the consistency of the Ar line with other reflection features should 
be checked  
using self-consistent reflection models including Ar and Ca 
transitions. 

We also tried to fit all the reflection features (the Compton hump, the iron 
edge and the iron line) with a self-consistent reflection model such as 
{\tt reflionx}, which we have modified with a high energy cutoff at the 
electron temperature of the Comptonizing corona to take into account the 
curvature of the Comptonization spectrum with respect to a simple power-law 
(used as illuminating spectrum in the reflionx model). The smearing of the 
reflection component has been taken into account multiplying it by the 
{\tt rdblur} component. In this case we had
to add a diskline to the model to fit the Ar line at 3 keV, since transitions
from Ar are not taken into account in the reflionx model. Anyway, all the
smearing parameters of the diskline used to fit the Ar line have been 
fixed to those used for the reflection component. In this case, we used 
a {\tt bbody} or a {\tt diskbb} to fit the soft thermal component and tried
to vary the iron abundance fixing this value to 1 or 2 times the cosmic
abundance. The best fit corresponds to a bbody component for the soft
thermal emission and to an iron abundance of 1 (see Table~2). 
We also tried to fit the reflection component with the self-consistent 
convolution model {\tt rfxconv}, which consistently takes into account the
curvature of the illuminating spectrum caused by the high-energy 
rollover at the electron temperature in the corona producing the 
primary Comptonized spectrum, or with the {\tt relxill} model, in which
the cutoff energy in the reflected spectrum is fixed at the electron temperature
of the primary Comptonized component, fitted with {\tt nthComp}. Note that
both rfxconv and relxill, as well as reflionx, all include Compton broadening
effects caused by Compton scattering in the surface layers of the accretion disc.
The results of these fits are reported in Table~3 and are perfectly consistent 
with those obtained with all the other reflection models that we tried.

\begin{table}
\caption{Comparison of the best-fit continuum and reflection parameters obtained
for the soft state (SS) as observed in the 60-ks {\xmm} observation and for the 
hard state (HS) observed by {\suzaku} (this paper). Continuum parameters for the 
SS observed by {\xmm} are taken by \citet{Egron.etal:13}, who use a similar model
for the continuum, while smearing parameters of the reflection component are taken
from \citet{Di_Salvo.etal:09} where these parameters are obtained with smaller 
uncertainties. $L_X$ is the X-ray luminosity extrapolated in the $0.1-150$ keV 
range for the SS, as reported by \citet{Egron.etal:13}, and in the $0.5-200$ keV 
range for the HS (this work). $L_{Edd}$ is the Eddington luminosity for a 
$1.4\;M_\odot$ NS, $L_{Edd} = 2.5 \times 10^{38}$ ergs s$^{-1}$ 
\citep{vanParadijs.etal:94}.
}
\vskip 0.5cm
\begin{tabular}{|lcc |}
\hline
Parameter & SS ({\xmm}) & HS ({\suzaku})  \\
\hline
$N_H$ ($\times 10^{22}$ cm$^{-2}$)  &  $2.08 \pm 0.02$  &  $2.11 \pm 0.04$ \\
bbody kT (keV)            & $0.56 \pm 0.01$ & $0.38 \pm 0.03$  \\
bbody $L_X$ ($L_{37} / D_{10}^2$)  &  $2.58 \pm 0.01$ &  $0.055 \pm 0.007$  \\
R$_{BB}$ (km)             & $33.3 \pm 1.2$  &  $10.6 \pm 1.8$  \\
nthComp $kT_{seed}$ (keV) & $1.30 \pm 0.02$  & $0.68 \pm 0.03$  \\
nthComp $kT_e$ (keV)      & $3.0 \pm 0.1$  & $47^{+20}_{-10}$  \\
$\xi$ (erg cm s$^{-1}$)   & $> 500$ & $< 13$  \\  
$Betor$                   & $-2.27 \pm 0.08$ & $-2.5 \pm 0.5$ \\  
$R_{in}$ ($G M / c^2$)    & $14 \pm 2$ & $17^{+4}_{-6}$ \\  
$R_{out}$ ($G M / c^2$)   & $3300^{+1500}_{-900}$ & $370^{+8000}_{-180}$ \\  
Incl (deg)                & $39 \pm 1$ & $43 \pm 5$ \\
$L_X/L_{Edd}$		  & $72\%$  &  $2.9\%$  \\
\hline
\end{tabular}
\end{table}

In order to check the stability of the best fit model with respect to 
the smearing parameters of the reflection component, we have let all
these parameters free to vary. The most uncertain of these parameters,
as expected, is the outer radius of the disk, for which we find only
loose constraints. Interestingly, all of the smearing parameters of the
best fit model of this observation are in good agreement with the 
smearing parameters already obtained with other instruments, 
e.g.\ {\xmm} and BeppoSAX \citep{Di_Salvo.etal:09, Egron.etal:13}, 
during a soft state. In order to facilitate the comparison, we 
report in Table~4 the best fit parameters of the reflection (modeled
by reflionx) and the relativistic smearing components obtained with {\xmm} 
during a soft state \citep[from][]{Di_Salvo.etal:09, Egron.etal:13} and 
obtained with {\suzaku} during a hard state (this paper). Although the 
uncertainties in the spectral parameters in the hard state are larger than 
in the soft state (because of the lower source flux in the hard state), 
we find a very good agreement in all the parameters. 
In particular, the inclination angles of the system we obtain in the two 
cases are compatible well within the 90\% c.l.\ uncertainty. The main 
difference in the reflection component between the hard and the soft state 
is in the ionization parameter $\xi$, which is much larger in the soft state 
than in the hard state, as expected because of the higher incident flux 
in the soft state. Also the continuum parameters are different; in 
particular, the temperatures of the soft components of the continuum 
(i.e.\ of the soft blackbody component and of the seed photons for the 
Comptonization) result higher in the soft state, while the electron 
temperature of the Comptonizing cloud results higher in the hard state,
in agreement with what is expected. 

We note that there is no clear indication of a receding inner accretion
disk radius in the hard state, corresponding to a luminosity of $\sim 3\%$
of the Eddington luminosity, with respect to the soft state, which was 
observed at about $70\%$ of L$_{Edd}$. On the 
contrary, the inner disk radius as inferred from the reflection component
is consistent to be the same in the two spectral states, at about 34 km
from the NS center. A similar indication comes from the inner radius of
the disk as inferred from the blackbody component, that we interpret as
the direct emission from the accretion disk. Both in the soft and in the
hard state the blackbody radius is a few tens of km, in agreement with 
the estimate we get from the reflection component. We caution the reader, 
however, that neither the color factor or the geometry of the system has 
been taken into account in this calculation. What is reported is just a 
zero-order estimation of the radius of the region (assumed to be spherical) 
emitting the blackbody component. In particular, the spectral hardening 
factor may depend on luminosity \citep[see e.g.][]{Merloni.etal:00}
explaining why the inner disk radius may appear larger at higher 
luminosities. 

This result is in agreement with what is found by \citet{Egron.etal:13}
who studied {\xmm}, BeppoSAX, and RXTE spectra of \1705 in the hard state and
in the soft state. In particular, in the hard state, the inner disk radius
as measured by the smearing of the reflection component resulted at 
$19-59$\, R$_g$, which is compatible with the inner disk radius derived in
the soft state ($13 \pm 3$\, R$_g$), while more uncertain results came
from the evaluation of the blackbody radius in the hard state. 
\citet{D_Ai.etal:10} also analyzed the same {\xmm} observation during a hard 
state used by \citet{Egron.etal:13}. These
authors discussed the possibility of a truncated disk in the hard state 
based on the best fit value of the inner disk radius as found from the Fe 
line width, which was about 30 R$_g$, that means about 60 km. However, in that 
case, the lack of broad--band coverage and the limited statistics, gave a large 
uncertainty on the inner disk radius, whose $90\%$ c.l.\ range was from 
6 to 90 R$_g$. Considering the large uncertainty on this measurement we 
cannot state that the result was in contrast with more recent results 
\citep[][and this work]{Egron.etal:13}. Note also that the best-fit blackbody 
radius reported by \citet{D_Ai.etal:10} has a value around $14 \pm 5$ km, 
indicating that the disk may be truncated quite close to the NS surface.
Similarly, \citet{Lin.etal:10} could not determine with high precision the 
inner radius of the disk using a diskline model for the Fe line in 
{\suzaku} and BeppoSAX spectra of \1705 taken during a hard state. 
Therefore, the inner radius of the disk was fixed to 6 R$_g$, and the fit 
results, such as Fe line flux and equivalent width, were not sensitive to 
this parameter. 

Similar results for the inner disk radius were obtained also in the case
of 4U 1728--34, the prototype of the atoll sources. The {\xmm} spectrum 
reported by \citet{Egron.etal:11} taken during a low-luminosity state of the
source (probably a hard state) showed a relatively broad iron line (Gaussian
$\sigma \sim 0.6$ keV), which was fitted to a series of models (diskline, 
relline, and reflionx) yielding in all the cases an inner disk radius 
between 12 and 22 R$_g$. In this case a blackbody component was not 
significantly detected. \citet{Cackett.etal:10} present a comprehensive, 
systematic analysis of {\suzaku} and {\xmm} spectra of 10 NS LMXBs, 
in order to study their Fe $K\alpha$ emission lines. In most cases
they find a narrow range of inner disk radii ($6-15$ R$_g$), implying 
that the accretion disk extends close to the NS surface over a range of 
luminosities. 

In this respect, it may be useful to compare these results to those obtained
for BH X-ray Binaries, since much work has been done to determine
the inner radius of the disk in these systems both from the iron line and
the reflection component and from the blackbody component 
\citep[e.g.][and references therein]{Done.etal:07, Reis.etal:10}. 
Also for these systems there is growing evidence that the disk may not be 
truncated far from the last stable orbit. The broad-band ($0.1-200$ keV) 
BeppoSAX spectrum of one of the best studied galactic BH candidates, 
Cygnus X--1, taken during a hard state showed evidence of a complex 
reflection component. In this spectral deconvolution the the inner radius 
of the disk, as inferred from the smeared reflection, is found between 
6 and 20 R$_g$ \citep{Di_Salvo.etal:01}. This result is in agreement with 
the results of \citet{Young.etal:01} who fitted ASCA, Ginga and EXOSAT data 
of Cygnus X--1 in both soft and hard spectral states to a model of an 
ionized accretion disk, whose spectrum is blurred by relativistic effects. 
They found that relativistic blurring provided a much better fit to the 
low/hard state and that data of both states were consistent with an 
ionized thin accretion disk with a reflected fraction of unity extending 
to the innermost stable circular orbit around the BH 
\citep[see, however,][for a different interpretation]{Barrio.etal:03}. 
Up to date, one of the strongest 
evidence of a truncated disk, based on Fe line profile measurements, in 
a BH hard state is that of GX 339-4 \citep{Tomsick.etal:09}. In that case,
from {\suzaku} and RXTE spectra, it was found that R$_{in}$ was a factor 
$>27$ higher than in the bright state when the luminosity was about at 
$0.14\%$ of the corresponding Eddington limit.

More recently \citet{Reis.etal:10} have analyzed a sample of 
stellar mass BHs, including Cygnus X--1, in the low-hard state, down to 
luminosities of $\sim 10^{-3}$ L$_{Edd}$, finding no clear evidence of 
a truncation of the inner disk at radii larger than 10\, R$_g$. 
Furthermore, the thermal-disk continuum yields colour 
temperatures consistent with the relation $L \propto T^4$, implying 
that the emitting surface is consistent with being constant with 
luminosity. A similar relation, $L \propto T^{3.2}$, seems also to hold 
in the case of \1705 at least for the soft state \citep{Lin.etal:10}. The
authors suggest that the deviation may be caused by a luminosity-dependent 
spectral hardening factor. On the other hand, the relatively low reflection 
amplitude ($\Omega/2\pi \simeq 0.34$) we find in the hard state of \1705 
fitting its spectrum to self-consistent reflection models, such as rfxconv,  
is compatible with a spherical geometry with the hot 
(spherical) corona inside an outer accretion disk. In this case,
the small inner radius of the disk we find (R$_{in} \sim 17$ R$_g$, 
corresponding to approximately $30-35$ km for a 1.4 M$_\odot$ NS) would
indicate a very compact hot corona filling the central part of the accretion 
disk, which in the case of a NS may be identified with a boundary layer
between the inner accretion disk and the NS surface. This might represent 
an important difference between BH and NS systems, since for a
BH, in the absence of a boundary layer, the inner disk should extend 
down to the last stable orbit. Note, however, that alternative explanations
for this ÔweakÕ reflection cannot be ruled out. This can be caused by a 
highly ionized inner disk \citep[e.g.][]{Ross.etal:99} or mildly 
relativistic outflow of the hot corona away from the disk \citep{Beloborodov:99}. 
In the latter case, it was shown that reflection fractions as low as $\sim 0.3$ 
can be obtained in the low-hard state without invoking a truncated disk. 

In summary, we have analyzed a deep (100 ks exposure time) ToO 
{\suzaku} observation of \1705 during a low-luminosity hard state (corresponding
to a luminosity of $\sim 3\%$ L$_{Edd}$). The broad-band spectrum shows a prominent 
Compton hump at hard energies, a Fe absorption edge and two relatively weak 
emission lines at $\sim 3$ and $\sim 6.4$ keV, identified with fluorescent 
emission from mildly ionized Ar and neutral Fe, respectively. We used all 
the available to date self-consistent reflection models to fit these 
reflection features finding in all cases best-fit parameters that are 
compatible with each other and consistent with those reported in literature 
for the soft state. 
In particular the inclination angles found from the reflection component for 
the hard and the soft state are perfectly compatible with each other.
It is worth noting that we obtain similar smearing parameters in the soft 
and in the hard state even if in the soft state the reflection component 
is dominated by the Fe line, while in the hard state other features are 
dominant. In fact the Fe line is the least important feature in the 
statistical sense (since the addition of other reflection features, 
such as the Compton hump plus the Fe edge, gave the most important 
improvement of the fit). 
We also find very similar inner disk radii in the hard and soft state, 
indicating that the inner disk rim does not change significantly at 
different spectral states down to a luminosity of $\sim 3\%$ the Eddington
limit.

\thanks{
We thank the unknown referee for useful comments which helped to improve the quality
of the paper.
The High-Energy Astrophysics Group of Palermo acknowledges support from the Fondo 
Finalizzato alla Ricerca (FFR) 2012/13, project N. 2012-ATE-0390, founded by the 
University of Palermo. This work was partially supported by the Regione Autonoma 
della Sardegna through POR-FSE Sardegna 2007-2013, L.R. 7/2007, Progetti di Ricerca 
di Base e Orientata, Project N. CRP-60529, and by the INAF/PRIN 2012-6.
EE acknowledges financial support from the Regione Autonoma della Sardegna through 
a research grant under the program CRP-25399 PO Sardegna FSE 2007-2013, L.R.\ 7/2007.
AP is supported by a Juan de la Cierva fellowship and acknowledges grants 
AYA2012-39303, SGR2009-811, and iLINK2011-0303.}

\bibliography{ms_1705_Suzaku_new}

\end{document}